\documentclass[]{spie}

\usepackage{graphicx}
\usepackage{array}
\usepackage{float}
\usepackage{simplemacros}
\usepackage[colorlinks=true, allcolors=blue]{hyperref}

\title{Scaling Datasets for Multi-Sensor, Multi-Agent, and Multi-Domain Learning in Autonomous Systems}

\author[a]{R. Spencer Hallyburton}
\author[a]{David Hunt}
\author[a]{Miroslav Pajic}
\affil[a]{Department of Electrical and Computer Engineering, Duke University, Durham, NC 27708 USA}

\authorinfo{E-mail: spencer.hallyburton@duke.edu; david.hunt@duke.edu; miroslav.pajic@duke.edu}

\begin{document}

\maketitle


\begin{abstract}
Existing datasets cannot support large-scale learning in multi-agent, multi-sensor, or multi-domain autonomy, where diversity and coordination are essential. We present a modular dataset generation pipeline that creates terabyte-scale, ground-truth-labeled data for ground, aerial, and infrastructure-based systems using the AVstack framework and CARLA simulator. Supporting single- and multi-agent configurations with flexible sensor suites, the pipeline enables controllable experimentation across challenging conditions. Representative perception and fusion studies show how generated data can support application-specific training and collaborative autonomy.
\end{abstract}


\section{Introduction}

Machine learning for autonomous systems remains constrained by the diversity and scale of available datasets. Real-world collections are expensive, narrowly scoped to specific platforms or sensing modalities, and difficult to label at the scale needed for supervised learning. These limits become more severe for collaborative autonomy, where evaluation depends not only on what one vehicle senses, but also on how heterogeneous sensors, agents, viewpoints, and communication patterns interact.

Benchmarks such as KITTI~\cite{2013kittidataset}, nuScenes~\cite{2020nuscenesdataset}, and Waymo~\cite{2020waymodataset} have enabled major advances in perception. More general computer-vision datasets such as PASCAL VOC~\cite{2010pascalvoc}, COCO~\cite{coco2014microsoft}, and Cityscapes~\cite{cordts2016cityscapes} similarly established supervised-learning tasks. However, these datasets are finite, fixed at collection time, and not designed for systematic sweeps over platform type, agent density, occlusion, communication, sensor degradation, or domain shift.

The limitations are concrete in practice. Many autonomy datasets are tied to a single vehicle platform, which fixes sensor heights, baselines, and mounting geometry for the entire dataset. Each camera or \lidar\ often has one fixed orientation, leaving no way to ask how a different pitch angle, mount height, roadside position, sensor rate, field of view, or overhead viewpoint changes performance. Many collections are dominated by passenger-vehicle traffic in ordinary weather, with limited coverage of heavy trucks, work zones, emergency maneuvers, adverse weather, night conditions, sensor degradation, construction layouts, rare occlusion patterns, or dense multi-agent interactions. These gaps do not make existing datasets unimportant, but they make them difficult to use for controlled studies of novel perception and collaboration scenarios.

Configurability is especially important for multi-sensor, multi-agent (\msma) autonomy. Vehicle-to-vehicle and vehicle-to-infrastructure collaboration are expected to be important components of future transportation systems~\cite{nhtsav2v}. OPV2V~\cite{2022openv2v} and related connected-autonomy datasets~\cite{2022cooperative} are important steps, but they remain tied to particular scenarios, agent models, and fusion assumptions. Many collaborative studies also treat data sharing as a small ego-centered problem, missing harder cases where infrastructure, aerial, and vehicle agents form a sensor network with different local histories, correlated estimates, and uneven visibility.

This paper presents a dataset-generation framework\footnote{Repository: \url{https://github.com/avstack-lab/carla-sandbox.git}.} built on \avstack~\cite{avstack,avstack-demo} and the \carla\ simulator~\cite{2017carla}. Rather than releasing a single static dataset, the framework turns dataset creation into a configurable experimental process. A user specifies agents, sensor payloads, scenario conditions, and logging hooks; the system records synchronized sensor data and object-state truth; postprocessing converts global truth into per-sensor labels through AVstack reference-frame and visibility tools. The same pipeline instantiates ground-vehicle datasets, elevated aerial camera networks, and infrastructure sensing deployments.

The contribution of this work is a framework for scalable dataset construction and downstream evaluation. The pipeline creates ground-truth-labeled \msma\ data across configurable ground, aerial, and infrastructure deployments; exposes the generated data through AVstack interfaces compatible with perception and dataset-conversion workflows; and supports repeatable studies of application-specific perception, collaborative fusion, security and trust, vision-language autonomy, and geometric scene understanding. The central systems lesson is that the value is not only the generated data, but also the ability to regenerate data under controlled changes to sensors, viewpoints, domains, agents, communication topology, and network assumptions.

\section{Dataset Generation Framework}

The framework separates dataset generation into four stages: scenario configuration, simulation and logging, postprocessing, and downstream formatting. This separation lets researchers vary agent placement, camera height, \lidar\ density, weather, or communication radius without rewriting the full autonomy stack.

\begin{figure}[t]
    \centering
    \includegraphics[width=0.94\linewidth]{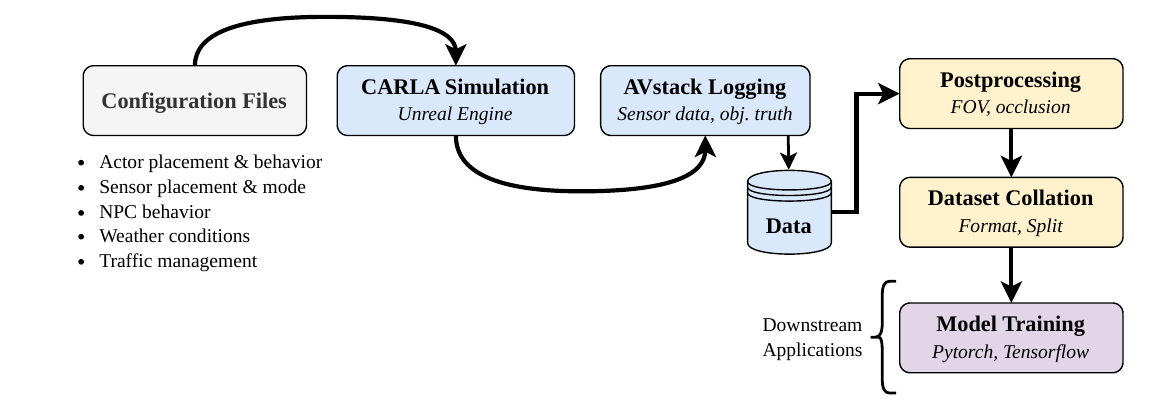}
    \caption{Dataset-generation workflow. Scenario configuration defines agents, sensors, maps, weather, and random seeds; CARLA execution produces synchronized sensor streams and object states; AVstack postprocessing converts global truth into per-sensor labels with calibration and visibility metadata; and downstream interfaces export generated data for AVstack and common training formats.}
    \label{fig:generation-pipeline}
\end{figure}

\subsection{Configuration and Simulation}

Scenarios are specified with human-readable configuration files that define actors, non-player characters (NPCs), and sensor suites. The CARLA sandbox passes configurations through to a runner that connects to a synchronous CARLA client, builds actors through the AVstack/CARLA registry, advances the world at a fixed simulation rate, and records data for a prescribed duration. Random seeds allow repeated collection with controlled variation.

Actor definitions support mobile ground or aerial vehicles and static elevated platforms. Sensor definitions include RGB camera, depth camera, semantic segmentation camera, \lidar, \radar, GNSS, and IMU models, all with configurable reference frames and CARLA attributes. Each sensor is represented as an AVstack object with calibration and reference-frame metadata. Logging hooks write sensor data and object states to run-specific folders, making a generated run a reproducible scene instance rather than a transient simulator session.

This configuration layer is the practical difference between simulation as a demonstration tool and simulation as a dataset generator. In a physical dataset, the distribution of other traffic participants is whatever happened during collection. In the generated setting, NPC behavior, traffic density, weather, map, sensor placement, and collection duration can be intentionally varied. That control is important for rare or safety-critical cases, such as occlusion-heavy traffic, abrupt maneuvers, degraded sensing, or adversarial communication patterns, and is consistent with the broader use of simulation platforms for autonomous-systems development and evaluation~\cite{2018airsim}.

\subsection{Reference Frames and Labels}

Multi-agent datasets require labels that are meaningful from each local sensor frame. A global object state alone is insufficient for camera or \lidar\ training because each sensor observes a different subset of the scene under a different calibration. Postprocessing uses AVstack reference frames to transform global object truth into each ego, infrastructure, or aerial sensor frame, then filters objects by field of view and visibility before writing labels.

The reference-frame problem is especially important outside homogeneous vehicle-to-vehicle settings. If every platform is a similar vehicle carrying sensors in similar poses, a global coordinate frame can hide much of the geometric complexity. Infrastructure and aerial sensing remove that simplifying assumption: sensors may be static, elevated, downward-facing, oblique, or arbitrarily oriented with respect to the road. AVstack's reference-frame chain keeps these transformations explicit so that labels, detections, tracks, and fused estimates can be moved between sensor, agent, and world frames without special-case code for each deployment.

Visibility is central to labeling. In CARLA, an object may exist in the world and project into a camera frustum or \lidar\ or \radar\ field of view while still being occluded by a building, vehicle, or other object; CARLA does not directly report visibility through occlusions. AVstack postprocessing estimates occlusion from available depth or point-cloud data through a geometry-based visibility algorithm and removes fully occluded or unknown objects from per-sensor labels. This produces consistent labels across heterogeneous sensors while preserving the fact that different agents observe different object subsets.

The postprocessor leverages depth and point cloud ray tracing to compare expected object depth with the depth obtained by a LiDAR or depth camera sensor measurement. Objects are then assigned visibility/occlusion states and filtered before annotations are written based on the occupancy of the filtered depth map. This avoids a common failure mode in simulator-derived datasets: treating all in-frustum objects as visible positives, even when they are hidden behind other scene geometry.

\subsection{Dataset Interfaces}

Generated data remain accessible through AVstack dataset managers, allowing the same downstream code to load generated CARLA data and common benchmarks. For camera-centric workflows, the sandbox includes a COCO conversion path that iterates over scenes, agents, sensors, and frames; links selected images into a COCO-style folder structure; and emits image and annotation metadata with 2D boxes projected from 3D truth. The broader AVstack integration also connects the generated data to model-training workflows, including camera-based detection, \lidar-based detection, depth estimation, and semantic segmentation pipelines.

The key point is that the framework is not a single dataset format frozen around one experiment. It is a configurable data-generation process with common abstractions for sensors, actors, references, labels, and postprocessing. That design lets the same pipeline support ground, aerial, and infrastructure sensing.

\section{Generated Dataset Domains}

The framework has been used in three complementary domains: mobile ground sensing, elevated aerial sensing, and infrastructure sensing. Each stresses a different weakness of fixed datasets. Ground-vehicle data resemble standard autonomous-vehicle benchmarks, aerial data introduce high-elevation overhead views with large spatial coverage, and infrastructure data introduce fixed roadside viewpoints with unique vantage points.

\subsection{Ground-Vehicle Sensing}

Ground agents are CARLA actors with vehicle-mounted sensors and CARLA-modeled autopilot behavior. Autonomy-derived behavior is possible but not explored here because the focus is dataset generation. The number of active and NPC agents in the scene can be scaled until CARLA processing bandwidth becomes the limiting factor. Any CARLA-defined sensor model can be attached to the platform and integrated in the data collection process, and custom sensor modes can be added through the same configuration pathway. Running scenarios with sensor attachments provides multi-agent ego-centric perception data similar to established autonomous-driving datasets, but with direct control over agent count, seed, duration, traffic composition, and sensor modality.

\begin{figure}[t]
    \centering
    \includegraphics[width=0.88\linewidth]{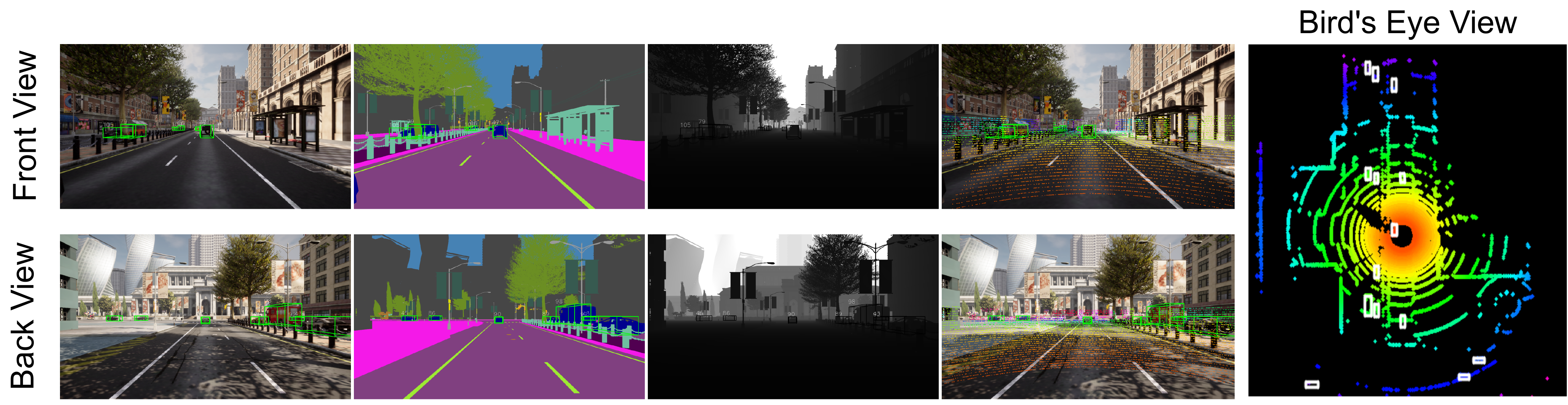}
    \caption{Representative multi-sensor data generated by the framework. The pipeline supports synchronized camera, depth, semantic, \lidar, and \radar\ sensing with object labels and calibration metadata.}
    \label{fig:dataset-sensor-visualization}
\end{figure}

Ground datasets can support full-scale learning as well as transfer learning in conjunction with real-world AV datasets. Figure~\ref{fig:dataset-sensor-visualization} illustrates sample multi-modality data collections from a run through the simulation. Ground datasets also expose the limits of ego-centric sensing: a single vehicle's field of view can be blocked by occlusion, constrained by sensor placement, or degraded by adverse conditions. This motivates infrastructure and aerial domains, where additional viewpoints can fill gaps in the ground vehicle's situational awareness.

Sample dataset collections demonstrate the scale that can be produced once dataset generation is treated as a repeatable pipeline. Two representative generated instances totaling nearly 1~TB of data were built through simulator scripting and auto-labeling in less than one hour: a multi-sensor vehicle dataset with multiple RGB, semantic-segmentation, depth, \lidar, and \radar\ sensors attached to an ego vehicle; and a multi-agent dataset with multiple ego-vehicles and infrastructure platforms carrying RGB, \lidar, and \radar\ sensors. Table~\ref{tab:dataset-statistics} summarizes the generated instances alongside state-of-the-art benchmarks. The important distinction is that the generated datasets are example outputs of a configurable generator, not fixed limits of the framework.

\begin{table}[H]
    \centering
    \caption{Representative dataset scale from prior generated instances compared with common benchmarks. The generated datasets are examples from short runs rather than fixed upper limits.}
    \label{tab:dataset-statistics}
    \small
    \begin{tabular}{>{\raggedright\arraybackslash}p{0.20\linewidth}>{\raggedright\arraybackslash}p{0.45\linewidth}>{\raggedright\arraybackslash}p{0.25\linewidth}}
        \hline
        Dataset & Sensing configuration & Scale summary \\
        \hline
        KITTI~\cite{2013kittidataset} & 1 \lidar\ and 4 RGB cameras & 7.5k images; 7.5k point clouds; 51k objects; 40 GB \\
        nuScenes~\cite{2020nuscenesdataset} & 1 \lidar, 6 RGB cameras, and 5 radars & 240k images; 40k \lidar\ frames; 280 GB \\
        OPV2V~\cite{2022openv2v} & Average of 3 connected agents, with \lidar\ and RGB sensing per agent & 40k images; 40k point clouds; 232k objects; 250 GB \\
        Generated multi-sensor (10 min.) & Multiple RGB, semantic, depth, \lidar, and \radar\ sensors on the ego vehicle & 200k images; 50k \lidar\ frames; 50k \radar\ frames; 3.1M objects; 500 GB \\
        Generated multi-agent (25 min.) & Four ego agents plus 5 infrastructure RGB and \lidar\ platforms & 300k images; 100k \lidar\ frames; 2M objects; 500 GB \\
        \hline
    \end{tabular}
\end{table}

\subsection{Aerial and Overhead Sensing}

Aerial configurations show that the same AVstack/CARLA machinery can generate overhead multi-sensor, multi-agent datasets. This is best positioned as an aerial or overhead sensing dataset rather than a vehicle-dynamics dataset. The configuration emphasizes sensor geometry unique to aerial vehicles: wide spatial coverage, reduced vehicle-to-vehicle occlusion, strong perspective shift relative to ego-vehicle datasets, and multi-camera overlap across the road network. These properties are exactly the kind of domain shift that fixed ground-vehicle datasets do not capture. The CARLA simulator is not presently designed to accurately model aerial vehicle dynamics, so the framework is limited in extensibility to kinematic analysis; however, perception-oriented insights are preserved. Compared with standard ego-vehicle datasets, fewer open-source datasets provide configurable aerial sensing with synchronized labels and multi-agent control.

\begin{figure}[t]
    \centering
    \begin{minipage}[t]{0.43\linewidth}
        \centering
        \includegraphics[width=\linewidth]{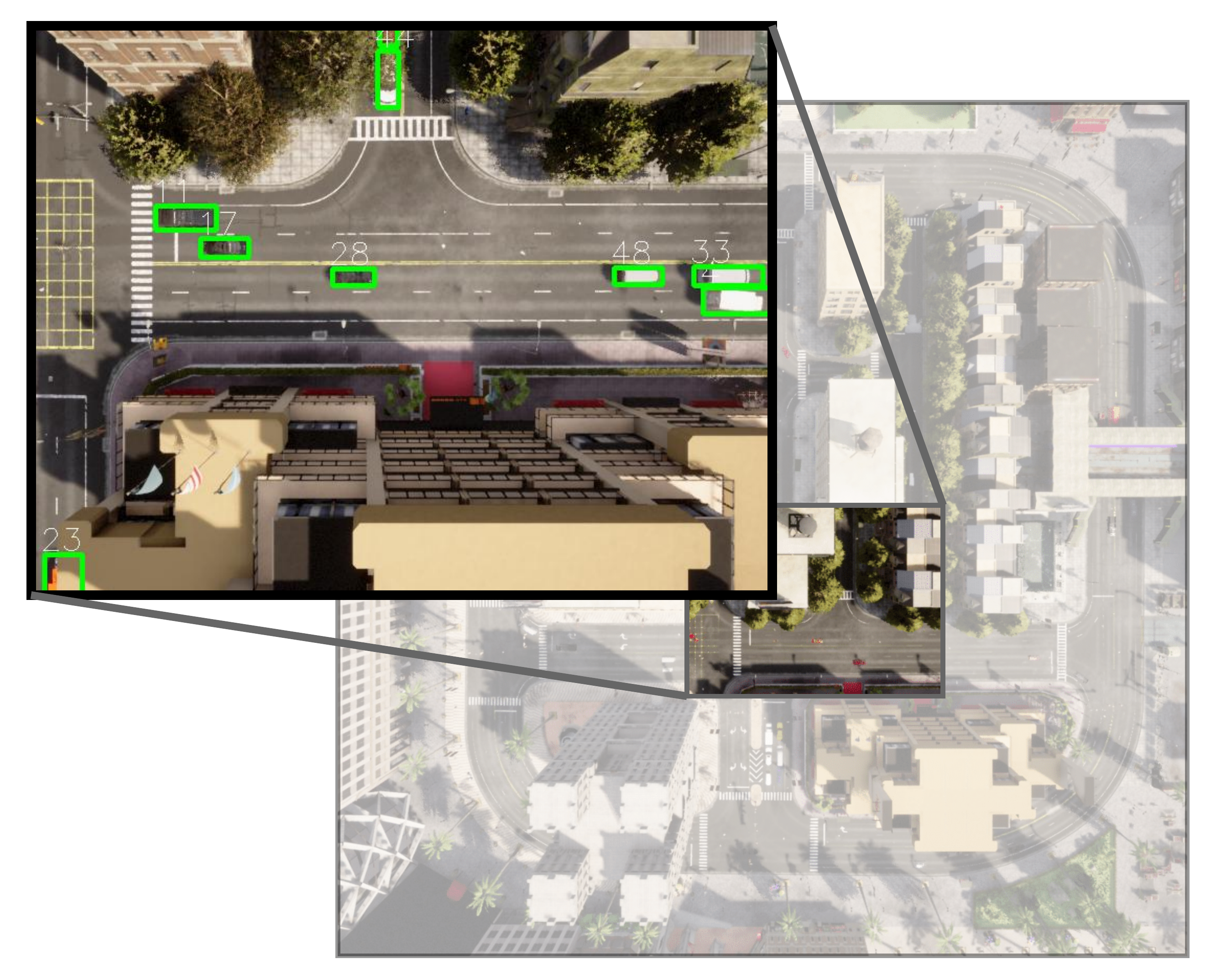}
        \vspace{2pt}
        {\small (a) Aerial capture and labels.}
    \end{minipage}
    \hfill
    \begin{minipage}[t]{0.43\linewidth}
        \centering
        \includegraphics[width=\linewidth]{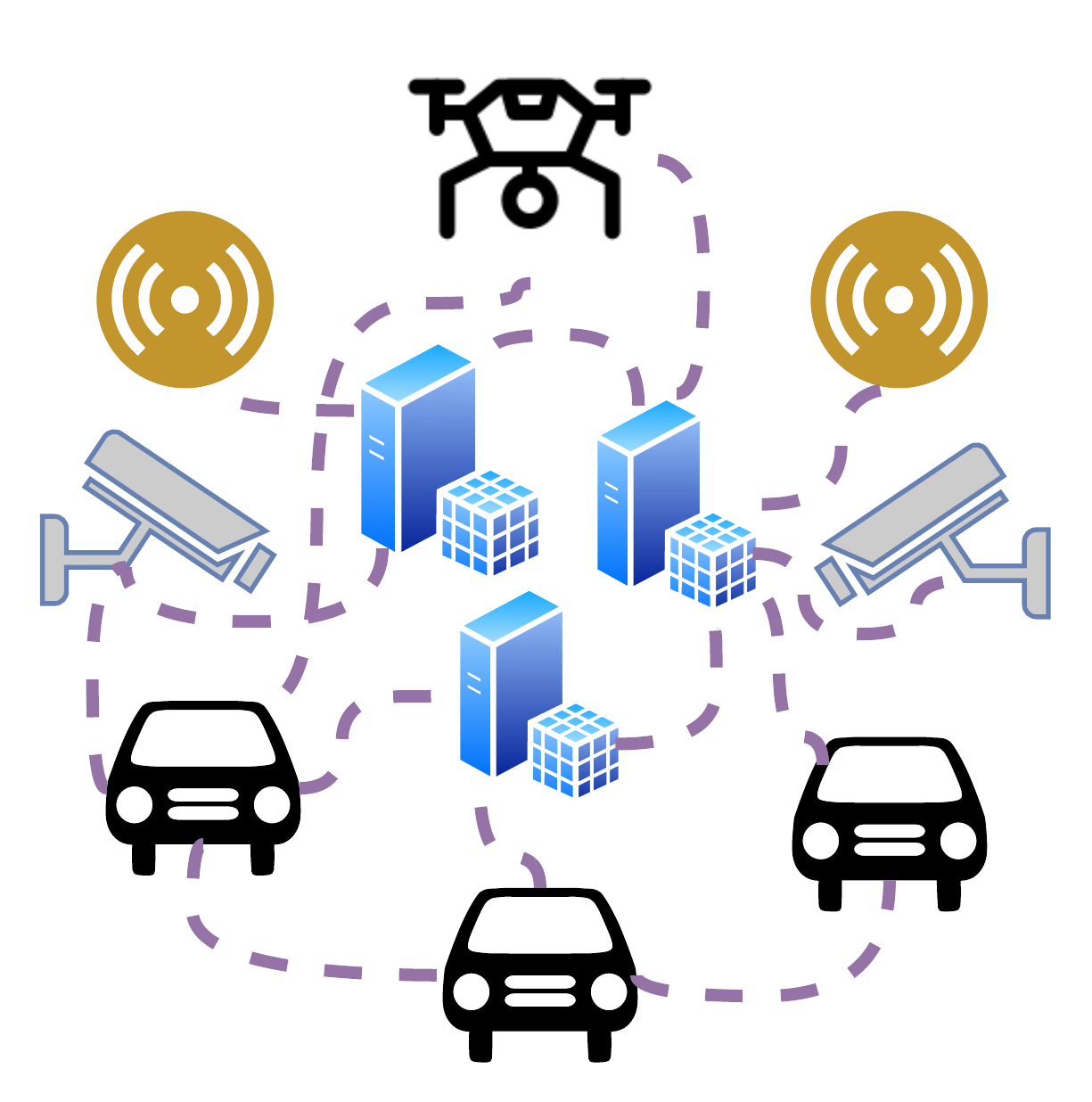}
        \vspace{2pt}
        {\small (b) Smart-city sensing topology.}
    \end{minipage}
    \caption{Aerial and smart-city dataset configurations. (a) A top-down CARLA scene with a zoomed overhead image capture from a drone-like aerial sensor; objects are automatically labeled with ground-truth bounding boxes through AVstack postprocessing. (b) Multi-agent smart-city sensing and communication topology, illustrating how vehicles, aerial platforms, and infrastructure sensors can provide complementary views for collaborative autonomy.}
    \label{fig:aerial-dataset-visualization}
\end{figure}

\subsection{Infrastructure Sensing}

\begin{figure}[t]
    \centering
    \includegraphics[width=0.86\linewidth]{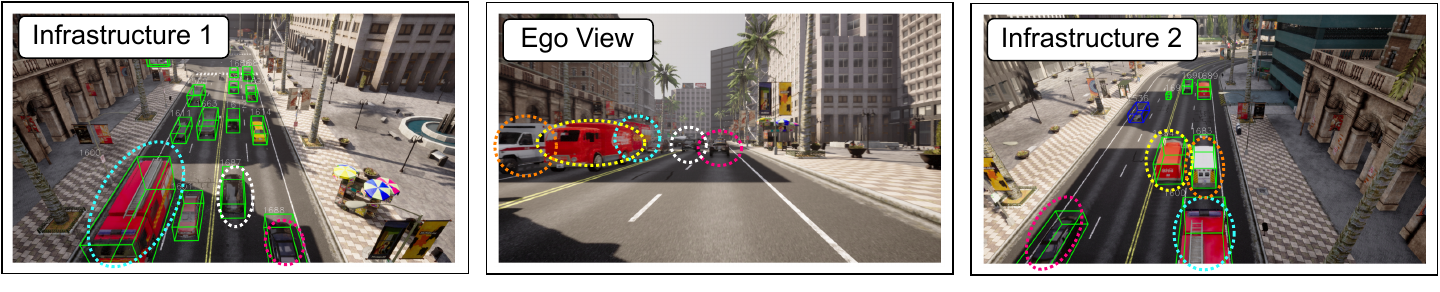}
    \caption{Infrastructure and ego viewpoints provide complementary observations of the same traffic scene. Multi-view geometry can reduce occlusion and improve situational awareness, but it also requires consistent reference-frame handling.}
    \label{fig:multi-agent-visualization}
\end{figure}

Infrastructure sensing provides non-ego viewpoints for collaborative autonomy. In a representative multi-agent configuration as described in Figure~\ref{fig:multi-agent-visualization}, an ego vehicle carries local RGB, \lidar, and \radar\ sensing while fixed infrastructure platforms provide additional observations. In smart-city deployments, roadside units mounted on traffic lights, poles, buildings, or gantries can observe intersections, crosswalks, merge zones, work zones, and blind approaches that may be occluded from any single vehicle. These persistent observers can also cover pedestrians, cyclists, trucks, and non-connected vehicles that cannot participate in vehicle-to-vehicle communication.

Unlike homogeneous vehicle-to-vehicle datasets, fixed infrastructure sensors can have arbitrary heights, orientations, fields of view, and communication pathways. AVstack's reference-frame chain is therefore central: it lets labels and detections from heterogeneous platforms be registered into common frames for downstream labeling and data fusion. Dataset generation standardizes these deployments because the same intersection can be regenerated under controlled changes to infrastructure height, orientation, sensor modality, traffic composition, weather, failure modes, and communication assumptions while retaining common labels for comparison.

\section{Downstream Applications}

Generated datasets are useful beyond inspection. They support controlled model training, cross-domain evaluation, and multi-agent algorithm studies. This section summarizes representative application areas.

\subsection{Application-Specific Perception}

Perception performance depends significantly on the viewpoint and domain of the training set compared to the test set. A detector trained on vehicle-mounted imagery may not transfer to a fixed infrastructure camera or elevated overhead camera, even when object classes are unchanged. Generated datasets make this failure mode testable because they provide labeled data from each sensing configuration.

One practical use of the multi-sensor, multi-agent dataset generation pipeline is producing large volumes of domain-specific multi-modal data for model training. Generated labels are exported into standard formats and can be used with existing training libraries such as MMDetection~\cite{chen2019mmdetection}, MMDetection3D~\cite{2020mmdet3d}, and representative perception pipelines such as Frustum PointNets~\cite{2018frustumpointnet}, SECOND~\cite{2018second}, and PointPillars~\cite{2019pointpillars}, or with custom model training pipelines that have AVstack plugins. This lets the dataset generator sit upstream of common 2D detection, 3D detection, depth, and semantic segmentation workflows rather than requiring a custom model stack.

We ran a controlled set of experiments with infrastructure and ground-vehicle data to highlight the utility of domain-specific adaptation; a complete description of the datasets, training setup, and model results appears in a companion analysis~\cite{hallyburton2023datasets}. In these experiments, object detection algorithms were trained and tested across vehicle and infrastructure datasets for both camera and \lidar\ sensing. As summarized in Table~\ref{tab:application-results}, camera models trained on vehicle-view data underperformed when evaluated on infrastructure imagery, while models trained on infrastructure data better matched the fixed-camera viewpoint. \lidar\ was less sensitive because point clouds preserve more 3D structure under sensor translation and rotation.

\begin{table}[H]
    \centering
    \caption{Illustrative application-specific perception results from the infrastructure study. Columns are [Train/Test] datasets. Values are class AP. The controlled comparison shows camera sensitivity to viewpoint shift in the generated data; Hallyburton and Pajic report the full experimental setup~\cite{hallyburton2023datasets}.}
    \label{tab:application-results}
    \begin{tabular}{llccc}
        \hline
        Modality & Class & Veh./Veh. & Infra./Infra. & Veh./Infra. \\
        \hline
        Camera & Truck & 0.33 & 1.00 & 0.00 \\
        Camera & Motorcycle & 0.00 & 1.00 & 0.00 \\
        Camera & Car & 0.51 & 0.67 & 0.00 \\
        Camera & Bicycle & 0.25 & 1.00 & 0.00 \\
        \lidar & Truck & 1.00 & 1.00 & 1.00 \\
        \lidar & Motorcycle & 1.00 & 1.00 & 1.00 \\
        \lidar & Car & 0.65 & 1.00 & 1.00 \\
        \lidar & Bicycle & 1.00 & 1.00 & 1.00 \\
        \hline
    \end{tabular}
\end{table}

\subsection{Collaborative Fusion}

Generated multi-agent data also support studies of how information should be shared across a sensor network. A common abstraction lets nearby agents send detections or tracks to an ego agent. This can improve recall when other agents see objects occluded from the ego view, but it also creates statistical challenges. If tracks have already been filtered over time or fused through neighboring agents, the incoming information may be correlated with the receiver's local estimate. The same data can also support multi-object tracking pipelines that combine 3D detection and temporal association across sensors~\cite{2020ab3dmot,2021eagermot}.

The infrastructure case study generated multi-agent data to compare local perception, detection-level fusion, and post-tracking distributed data fusion. Fusion at perception shares raw sensor data or low-level features; it can be accurate, but it is bandwidth intensive and difficult to secure, especially in collective-perception settings that must account for V2X misbehavior~\cite{ansari2021v2x}. Fusion at tracking shares semantic detections or tracks, which is cheaper, but it can incorrectly treat correlated inputs as independent measurements. Post-tracking distributed data fusion instead treats collaboration as a decentralized sensor-network problem and fuses track estimates conservatively, for example using covariance intersection~\cite{1994ddfframework,2017ddfwithCI}.

The experiment modeled an ego agent receiving information from an infrastructure-aided network under three correlation assumptions: no correlation, minor correlation, and major correlation. Collaboration improved situational awareness, but naive fusion became fragile under correlated network assumptions. Conservative track-to-track fusion was better aligned with the decentralized sensor-network setting. Table~\ref{tab:collaborative-results} highlights the improvements under collaborative tracking as well as the challenges introduced by correlated sensor networks.

\begin{table}[H]
    \centering
    \caption{Collaborative infrastructure fusion results. Values are mAP under increasing correlation assumptions.}
    \label{tab:collaborative-results}
    \begin{tabular}{lccc}
        \hline
        Assumption & Local only & Naive Fusion & Correlation-Aware Fusion \\
        \hline
        No correlations & 0.60 & 0.86 & 0.90 \\
        Minor correlations & 0.60 & 0.74 & 0.89 \\
        Major correlations & 0.60 & 0.62 & 0.86 \\
        \hline
    \end{tabular}
\end{table}

\subsection{Security and Trust Studies}

The same generated datasets can serve as testbeds for security and trust evaluation. In a ROS2-based multi-agent security testbed, generated multi-agent data supported repeatable studies of compromised collaborators, centralized fusion vulnerabilities, and adversarial false-positive and false-negative behavior~\cite{hallyburton2024multi}. More broadly, trust-aware multi-agent decision making and resilient autonomy under malicious participants have been studied in networked and robotic systems~\cite{liu2004dynamic,zhu2004computing,cavorsi2024exploiting}. Subsequent trust-estimation studies used related generated scenarios to evaluate Bayesian trust updates, trust pseudomeasurements, and trust-weighted sensor fusion under compromised-agent threat models~\cite{hallyburton2024bayesian,hallyburton2025security}. The aerial dataset generation process extends this use case to distributed overhead sensing, where agents perform local tracking, distance-limited communication, covariance-intersection fusion, and trust-informed filtering in adversarial settings~\cite{hallyburton2025trust}. These studies illustrate why controllable dataset generation matters: security experiments require repeated, instrumented variation over agent count, communication topology, sensing geometry, attack capability, and random seed.

\subsection{Advanced Algorithm Studies}

The dataset-generation process has supported algorithm studies beyond conventional perception and tracking. It can support end-to-end and multi-modal driving studies, including imitation- and reinforcement-learning pipelines in simulation~\cite{2020learningbycheating,2020marlcarla}, multi-modal driving architectures~\cite{2021transfuser}, and trajectory-prediction models that require heterogeneous scene context and agent histories~\cite{2020trajectron++}. It has also been used in vision-language-action models to create multi-sensor, multi-agent data for training and evaluating VLM/VLA-based driving assistants that reason about scene context, object motion, and trajectory planning~\cite{liu2025llavida}. For geometric scene understanding, generated data have supported field-of-view estimation studies, where models infer visible regions and uncertainty from perception data rather than assuming known sensor coverage~\cite{hallyburton2025probabilistic}. These studies capture the same benefit as the perception and security experiments: the framework can generate labeled data for tasks whose ground truth is difficult to obtain from ordinary driving logs.

Across these applications, the common pattern is that dataset generation acts as an experimental control surface. Rather than treating a dataset as a fixed benchmark, the framework lets researchers vary viewpoint, sensor placement, agent count, communication topology, and sensor domain while preserving synchronized labels and common dataset interfaces. This makes the generated data useful both for training models and for testing system-level autonomy assumptions that are difficult to isolate in real-world collections.

\section{Conclusion}

We presented a framework for generating configurable datasets for multi-sensor, multi-agent, and multi-domain autonomy. By combining \avstack's sensor, reference-frame, logging, dataset, and model interfaces with the \carla\ simulator, the pipeline creates labeled data from ground vehicles, aerial platforms, and infrastructure sensors. The central benefits are volume and control: researchers can vary platforms, sensors, viewpoints, agent counts, communication topology, weather, and scenario seeds while retaining synchronized data and per-sensor labels. The same generated data has supported perception, fusion, security and trust, vision-language modeling, and geometric scene-understanding studies, illustrating how configurable dataset generation can make difficult autonomy scenarios easier to isolate and study.

\acknowledgments

This work is sponsored in part by the ONR under agreement N00014-23-1-2206, AFOSR under the award number FA9550-19-1-0169, and by the NSF under NAIAD Award 2332744 as well as the National AI Institute for Edge Computing Leveraging Next Generation Wireless Networks, Grant CNS-2112562.

\bibliographystyle{spiebib}
\bibliography{references}


\end{document}